\pdfoutput=1

\documentclass[letterpaper,twocolumn,10pt]{article}
\usepackage{usenix2019,epsfig}

\usepackage{booktabs}       
\usepackage{changepage}     
\usepackage{comment}        
\usepackage{enumitem}       
\usepackage{epsfig}         
\usepackage{etoolbox}       
\usepackage{paralist}       
\usepackage{subcaption}     
\usepackage{titling}        
\usepackage[normalem]{ulem}   
\usepackage{upgreek}        
\usepackage{url}            
\usepackage{xcolor}         

\newtoggle{showmarks}
\toggletrue{showmarks}

\definecolor{darkgreen}{rgb}{0,0.7,0}
\definecolor{darkred}{rgb}{0.7,0,0}
\definecolor{orange}{rgb}{1,0.4,0}

\iftoggle{showmarks}{
\newcommand\red[1]{\textcolor{red}{#1}}
\newcommand\redstrike[1]{\red{\sout{#1}}}
\newcommand\green[1]{\textcolor{darkgreen}{#1}}
\newcommand\greenstrike[1]{\green{\sout{#1}}}
\newcommand\greenstrikealt[1]{\green{\sout{#1}}}
\newcommand\orange[1]{\textcolor{orange}{#1}}
\newcommand\orangestrike[1]{\orange{\sout{#1}}}
\newcommand\blue[1]{\textcolor{blue}{#1}}
\newcommand\bluestrike[1]{\blue{\sout{#1}}}
\newcommand\todo[1]{\textcolor{blue}{(TODO: #1)}}
}{
\newcommand\red[1]{#1}
\newcommand\redstrike[1]{\unskip}
\newcommand\green[1]{#1}
\newcommand\greenstrike[1]{\unskip}
\newcommand\greenstrikealt[1]{}
\newcommand\orange[1]{#1}
\newcommand\orangestrike[1]{\unskip}
\newcommand\blue[1]{\unskip}
\newcommand\bluestrike[1]{\unskip}
\newcommand\MA[1]{\unskip}
\newcommand\todo[1]{\unskip}
}

\usepackage{caption}
\usepackage{titlesec}
\titlespacing*{\section}{0pt}{7pt plus 3pt minus 3pt}{3pt plus 3pt minus 2pt}
\titlespacing*{\subsection}{0pt}{4pt plus 3pt minus 2pt}{1pt plus 3pt minus 1pt}
\titlespacing*{\subsubsection}{0pt}{4pt plus 3pt minus 2pt}{0pt plus 3pt minus 1pt}
\titleformat{\section}{\large\bfseries}{\thesection}{1em}{}
\titleformat{\subsection}{\normalfont\bfseries}{\thesubsection}{1em}{}

\renewcommand\smallskip{\vspace{2pt}}

\newcommand\figSqueeze{-3.0mm}

\begin{document}

\setlength{\droptitle}{-0.6in}

\title{\Large \bf It's Time to Replace TCP in the Datacenter}

\author{
{\rm John Ousterhout}\\
Stanford University
}

\date{January 18, 2023\\
\begin{minipage}{18cm}
\vspace{0.2in}
\normalsize
\emph{This position paper has been updated since its original publication
in October of 2022 in order to correct errors and add clarification.
Updates are in italics; none of the original text has been modified.
The paper has triggered discussion and dissent; for pointers to
comments on the paper, see the Homa Wiki:
\url{https://homa-transport.atlassian.net/wiki/spaces/HOMA/overview\#replaceTcp}.}
\end{minipage}
}


\maketitle

\begin{abstract}
In spite of its long and successful history, TCP is a poor
transport protocol for modern datacenters.
Every significant element of TCP, from its stream
orientation to its expectation of in-order packet delivery, is
wrong for the datacenter.
It is time to recognize that TCP's problems are too fundamental and
interrelated to be fixed; the only way to harness the full performance
potential of modern networks is to introduce a new
transport protocol into the datacenter.
Homa demonstrates that it is possible to create a
transport protocol that avoids all of TCP's problems.
Although Homa is not API-compatible with TCP, it should be possible
to bring it into widespread usage by integrating it with RPC frameworks.
\end{abstract}

\section{Introduction}

The TCP transport protocol~\cite{tcp-rfc} has proven to be phenomenally
successful and adaptable. At the time of TCP's design in the late 1970's,
there were only about 100 hosts attached to the existing ARPANET, and
network links had speeds of tens of kilobits/second.
Over the decades since then, the Internet has grown to billions of hosts
and link speeds of 100 Gbit/second or more are commonplace, yet TCP
continues to serve as the workhorse transport protocol for almost all
applications. It is an extraordinary engineering achievement to have
designed a mechanism that could survive such radical changes in underlying
technology.

However, datacenter computing creates unprecedented challenges for TCP.
The datacenter environment, with millions of cores in close proximity and
individual applications harnessing thousands of machines that interact on
microsecond timescales, could not have been envisioned by the designers
of TCP, and TCP does not perform well in this environment.
TCP is still the protocol of choice for most datacenter
applications, but it introduces overheads on many levels, which limit
application-level performance.
For example, it is well-known that TCP suffers from high tail latency for
short messages under mixed workloads~\cite{dctcp}.
TCP is a major contributor to the
``datacenter tax''~\cite{killerMicros, dctax}, a collection of low-level
overheads that consume a significant fraction of all processor cycles
in datacenters.

This position paper argues that TCP's challenges in the datacenter are
insurmountable.
Section \ref{sec:tcpProblems} discusses each of the major design
decisions in TCP and demonstrates
that every one of them is wrong for the datacenter, with significant
negative consequences.
Some of these problems have been discussed in the past, but it is instructive
to see them all together in one place.
TCP's problems impact systems at multiple levels, including the network,
kernel software, and applications.
One example is load balancing, which is essential in datacenters
in order to process high loads concurrently. Load balancing did not exist
at the time TCP was designed, and TCP interferes with load balancing
both in the network and in software.

Section \ref{sec:beyondRepair} argues that TCP cannot be fixed in an
evolutionary fashion; there are too many problems and too many interlocking
design decisions.
Instead, we must find a way to introduce a radically different
transport protocol into the datacenter.
Section \ref{sec:homa} discusses what a good
transport protocol for datacenters should look like, using
Homa~\cite{homa, homaLinux} as an example.
Homa was designed in a clean-slate fashion to meet the needs of
datacenter computing, and virtually every one of its major design decisions
was made differently than for TCP.
As a result, some problems, such as congestion in the network core fabric,
are eliminated entirely.
Other problems, such as congestion control and load balancing, become
much easier to address.
Homa demonstrates that it is possible to solve all of TCP's problems.

Complete replacement of TCP is unlikely anytime soon, due
to its deeply entrenched status, but TCP can be displaced for many
applications by integrating Homa into a small number of existing RPC
frameworks such as gRPC~\cite{grpc}.
With this approach, Homa's incompatible API will be visible only
to framework developers and applications should be able to switch to
Homa relatively easily.

\section{Requirements}

Before discussing the problems with TCP, let us first review the
challenges that must be addressed by any transport protocol for
datacenters.

\smallskip
\noindent{\bf Reliable delivery.} The protocol must deliver data
reliably from one host to another, in spite of transient failures
in the network.

\smallskip
\noindent{\bf Low latency.} Modern networking hardware enables round-trip
times of a few microseconds for short messages. The transport protocol must
not add significantly to this latency, so that applications experience
latencies close to the hardware limit. The transport protocol must also
support low latency at the tail, even under relatively high network loads with
a mix of traffic.
Tail latency is particularly challenging for transport protocols; nonetheless,
it should be possible to achieve tail latencies
for short messages within a factor of 2--3x of the best-case latency
\cite{homa}. 

\smallskip
\noindent{\bf High throughput.}
The transport protocol must support high throughput in two different ways.
Traditionally, the term ``throughput'' has referred to \emph{data throughput}:
delivering large
amounts of data in a single message or stream. This kind of
throughput is still important. In addition, datacenter applications
require high \emph{message throughput}: the ability to
send large numbers of small messages quickly for communication
patterns such as broadcast and shuffle~\cite{millisort}.
Message throughput has historically not received much attention, but
it is essential in datacenters.

In order to meet the above requirements, the transport protocol must
also deal with the following problems:

\smallskip
\noindent{\bf Congestion control.}
In order to provide low latency, the transport protocol must limit the
buildup of packets in network queues. Packet queuing can potentially
occur both at the edge (the links connecting hosts to top-of-rack switches)
and in the network core; each of these forms of congestion creates
distinct problems.

\smallskip
\noindent{\bf Efficient load balancing across server cores.}
For more than a decade, network speeds have been increasing rapidly
while processor clock rates have remained nearly constant. Thus it
is no longer possible for a single core to keep up with a single
network link; both incoming and outgoing load must be distributed across
multiple cores. This is
true at multiple levels. At the application level, high-throughput
services must run on many cores and divide their work among
the cores. At the transport layer, a single core cannot keep up
with a high speed link, especially with short messages. Load balancing
impacts transport protocols in two ways. First, it
can introduce overheads (e.g. the use of multiple
cores causes additional cache misses for coherence). Second, load
balancing can lead to \emph{hot spots},
where load is unevenly distributed across cores; this is a form of congestion
at the software level. Load balancing overheads are now one of the primary
sources of tail latency~\cite{homaLinux}, and they are impacted by
the design of the transport protocol.

\smallskip
\noindent{\bf NIC offload.}
There is increasing evidence that software-based transport protocols
no longer make sense; they simply cannot provide high performance at an
acceptable cost. For example:
\begin{compactitem}
\item The best software protocol implementations have end-to-end
latency more than 3x as high as implementations where applications communicate
directly with the NIC via kernel bypass.
\item Software implementations give up a factor of 5--10x in small message
throughput, compared with NIC-offloaded implementations.
\item Driving a 100 Gbps network at 80\% utilization in both directions
consumes 10--20 cores just in the networking
stack~\cite{snap, homaLinux}. This is not a cost-effective use of resources.
\end{compactitem}
Thus, in the future, transport protocols will need to move to
special-purpose NIC hardware. The transport protocol must not have
features that preclude hardware implementation. Note that NIC-based
transports will not eliminate software load balancing as an issue:
even if the transport is in hardware, application software will still
be spread across multiple cores.
\section{Everything about TCP is wrong}
\label{sec:tcpProblems}

This section discusses five key properties of TCP, which cover
almost all of its design:
\begin{compactitem}
\item Stream orientation
\item Connection orientation
\item Bandwidth sharing (``fair'' scheduling)
\item Sender-driven congestion control
\item In-order packet delivery
\end{compactitem}
Each of these properties represents the wrong decision for a datacenter
transport, and each of these decisions has serious negative consequences.

\subsection{Stream orientation}
The data model for TCP is a stream of bytes. However, this is not the
right data model for most datacenter applications. Datacenter
applications typically exchange discrete \emph{messages} to implement
remote procedure
calls. When messages are serialized in a TCP stream, TCP has no
knowledge about message boundaries. This means that when an application
reads from a stream, there is no guarantee that it will receive a complete
message; it could receive less than a full message, or parts of several
messages.
TCP-based applications must mark message boundaries when they serialize
messages (e.g., by prefixing each message with its length), and
they must use this information to reassemble messages on receipt.
This introduces extra complexity and overheads, such as maintaining
state for partially-received messages.

The streaming model is disastrous for software load balancing.
Consider an application that uses a collection of threads to serve
requests arriving across a collection of streams. Ideally, all of the
threads would wait for incoming messages on any of the streams, with
messages distributed across the threads.
However, with a byte stream model there is no guarantee that a
read operation returns an entire message. If multiple threads
both read from a stream, it is possible that parts of a single message
might be received by different threads.
In principle it might be possible for the threads to coordinate
and reassemble the entire message in one of the threads, but this
is too expensive to be practical.

Instead, TCP applications must use one of two inferior forms of load
balancing, in which each stream is owned by a single thread.
The first approach, used
by memcached~\cite{memcached}, is to divide a collection of streams
statically among the threads, where each thread handles all of the
requests arriving on its streams. This approach is prone to hot spots, where
one thread receives a disproportionate share of incoming requests.
The second approach, used in RAMCloud~\cite{ramcloud-tocs}, dedicates
one thread to read all incoming messages from all streams and then
dispatch messages to other threads for service. This allows much better
load balancing across worker threads, but the dispatcher thread becomes
a throughput bottleneck. Furthermore, the need for each request to pass
through two separate threads adds significant software overhead and
increases latency. Thus, the dispatcher thread approach
is effective only if request service times are relatively long.

The fundamental problem with streaming is that the units in which
data is received (ranges of bytes) do not correspond to
dispatchable units of work (messages).
There is no point in waking up a thread to receive part of a message;
it will not be able to process the message until it receives the entire message.
And, if a thread receives multiple messages in a single read
operation, it can only process one of them at a time; it would be better
for each message to be dispatched to a different thread so the messages can
be processed concurrently.

Streaming's negative impact on load balancing will carry over into
a future world where transport processing is offloaded to the NIC.
In this world, the NIC should perform load balancing, dispatching
incoming requests across a collection of application threads via
kernel bypass.
However, this will not be possible, since information about message
boundaries is application-specific and unknown to the transport
layer. Applications will still have to use one of the approaches 
described above, each of which impacts latency and/or throughput.

Streaming has an additional impact on tail latency because it induces
head-of-line blocking. Messages sent on a single stream must be
received in order; this means that a short message can be delayed
behind a long message in the same stream.
We observed this phenomenon in RAMCloud, where small time-sensitive
replication requests
from one server to another could be delayed by long background requests
for log compaction, resulting in a 50x increase in write
latency~\cite{ramcloud-tocs}.

Finally, the reliability guarantees provided by streaming are not the
right ones for applications.
Applications want \emph{round trip} guarantees.
A client application wants an assurance that its
request will be delivered and processed, and that it will receive a
response; if any of these fails, the client would like an error notification.
However, a stream guarantees only best-effort delivery of data
in one direction. The client will receive no notification
if the server does not send a response, and
under some conditions there will be no notification if the server machine
crashes.
As a result, clients must implement their own end-to-end timeout mechanisms,
even though TCP already has timers of its own.
These mechanisms introduce additional overheads.

\subsection{Connection orientation}
TCP requires long-lived connection state for each peer that
an application communicates with.
Connections are undesirable in datacenter environments because
applications can have hundreds or thousands of them, resulting
in high overheads in space and/or time.
For example, the Linux kernel keeps about 2000 bytes of state for each
TCP socket, excluding packet buffers; additional state is required
at application level.

Facebook found the memory demands for a separate connection
between each application thread and each server ``prohibitively
expensive''~\cite{scalingMemcached}. To reduce these overheads,
application threads communicate through a collection of proxy threads that
manage connections to all the servers. This allows a single connection
for each server to be shared across all the application threads on that
host, but it adds overhead for communicating through the proxies.
To reduce the proxy overheads, Facebook uses UDP instead of TCP for
requests that can tolerate UDP's unreliability, but this sacrifices
congestion control.

The overheads for connection state are also problematic when offloading the
transport to the NIC, due to limited resources on the NIC chip.
This problem is well known in the Infiniband
community~\cite{farm-nsdi, efficientRdma, rdmaDesignGuidelines}.
For many years, RDMA NICs could cache the state for only a few hundred
connections; if the number of active connections exceeded the cache size,
information had to be shuffled between host memory and the NIC, with a
considerable loss in performance.

Another problem with connections is that they require a setup phase before
any data can be transmitted. In TCP the setup phase has a nontrivial cost,
since it requires an additional round-trip between the hosts.
Traditionally, connections have been long-lived, so the setup cost
can be amortized across a large number of requests.
However, in new serverless environments applications
have very short lifetimes, so it is harder to amortize the cost of
connection setup.

It seems to be an article of faith in the networking community that
connections are required in order to achieve desirable properties
such as reliable delivery and congestion control, but connections carry
a high cost and Section~\ref{sec:homa} will show that it is possible
to achieve these properties without connections.

\subsection{Bandwidth sharing}
In TCP, when a host's link is overloaded (either for incoming or outgoing
traffic), TCP attempts to share the available bandwidth equally among
the active connections. This approach is also referred to as
``fair scheduling''.

Unfortunately, scheduling disciplines like this are well known to perform
poorly under load. When receiving several large messages, bandwidth
sharing causes all of them to finish slowly. Run-to-completion
approaches such as SRPT (Shortest Remaining Processing Time) provide
better overall response time because they dedicate all of the available
resources to a single task at a time, ensuring that it finishes quickly.
It is difficult implement run-to-completion with TCP because TCP has no
information about message boundaries; thus, it does not know when a task is
``complete''.

\begin{figure}
\begin{center}
\includegraphics[scale=0.52]{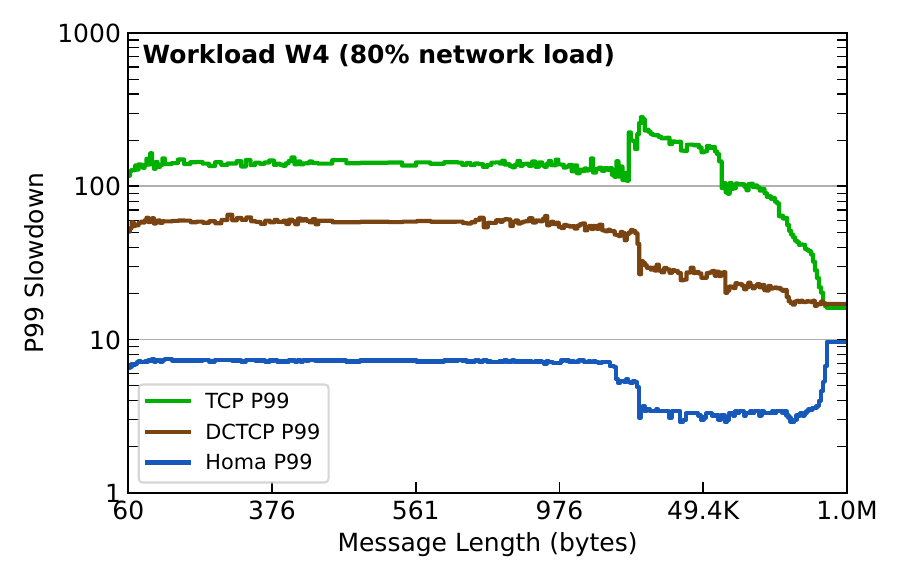}
\caption{99th percentile slowdown as a function of message length for
Linux kernel implementations of TCP, DCTCP, and Homa, running on
a 40-node CloudLab cluster with 25 Gbps network links running at 80\%
average utilization (see \cite{homaLinux} for details).
The workload is based on a message size distribution measured on
a Hadoop cluster at Facebook.
Slowdown is the round-trip trip time for a message
on a loaded cluster, divided by the time for Homa messages of the same length
in an unloaded system.
}
\label{fig:p99_w4}
\vspace{\figSqueeze{}}
\end{center}
\end{figure}

Furthermore, in spite of the name ``fair scheduling'', TCP's approach
discriminates heavily against short messages.
Figure \ref{fig:p99_w4} shows how round-trip latencies for messages
of different sizes slow down when running on a heavily loaded network,
compared to messages of the same size on an unloaded network.
With TCP, short messages suffer a slowdown almost 10x worse than the
longest messages. DCTCP reduces the gap somewhat, but short messages
still suffer 3x worse treatment than long ones.
Short message latency is critical in datacenter environments, so this
discrimination is problematic.

\subsection{Sender-driven congestion control}
\label{sec:tcpCongestion}
TCP drives congestion control from senders, which voluntarily slow their
rate of packet transmission when they detect congestion.
Senders have no first-hand knowledge of congestion, which can happen either
in the core fabric or at edge links between top-of-rack switches and receivers,
so they rely on congestion signals related to buffer occupancy.
In the worst case, switch queues overflow and packets are dropped,
leading to timeouts.
More commonly, switches generate ECN notifications when queue lengths
reach a certain threshold~\cite{dcqcn}, or senders detect increases in
round-trip times due to queueing~\cite{timely, swift}; some newer approaches
use programmable switches to generate more precise information such
as exact queue lengths~\cite{hpcc, powertcp}.
Senders then use this information to back off on packet transmission.

Congestion control in TCP is hobbled by two limitations.
First, congestion can only be detected when there is buffer occupancy;
this virtually guarantees some packet queueing when the network
is loaded. Second, TCP does not take advantage of the priority queues
in modern network switches. Thus, all packets are
treated equally and queues generated by long messages (where throughput
matters more than latency) will cause delays for short messages.

These limitations lead to a ``pick your poison'' dilemma where
it is difficult to simultaneously optimize both latency and
throughput.
The only way to ensure low latency for short messages is to keep queue
lengths near zero in the network.
However, this risks buffer under-runs, where links are idle even though
there is traffic that could use them; this reduces throughput for
long messages.
The only way to keep links fully utilized in the face of traffic
fluctuations is to allow buffers to accumulate in the steady state, but
this causes delays for short messages.

Furthermore, it takes about 1 RTT for a sender to find out about
traffic changes, so senders must make decisions based on
out-of-date information. As messages get shorter and networks get faster,
more and more messages will complete in less than 1 RTT, which makes
the information received by senders less and less reliable.

Congestion control has been studied extensively, both for TCP and
for other streaming approaches such as RDMA. These efforts have
resulted in considerable improvements, but it is unlikely that the
latency vs. throughput dilemma can be completely resolved without
breaking some of TCP's fundamental assumptions.

\subsection{In-order packet delivery}
TCP assumes that packets will arrive at a receiver in the same order
they were transmitted by the sender, and it assumes that out-of-order
arrivals indicate packet drops. This severely restricts load balancing,
leading to hot spots in both hardware and software, and
consequently high tail latency.

In datacenter networks, the most effective way to perform load balancing
is to perform \emph{packet spraying}, where each packet is independently routed
through the switching fabric to balance loads on links.
However, packet spraying cannot be used with TCP since it could change
the order in which packets arrive at their destination.
Instead, TCP networks must use \emph{flow-consistent routing},
where all of the packets
from a given connection follow the same path through the network fabric.
Flow-consistent routing ensures in-order packet delivery, but it
virtually guarantees that there will be overloaded links in the network
core, even when the overall network load is low.
All that is needed for congestion is for two large flows to hash
to the same intermediate link; this hot spot will persist
for the life of the flows and cause delays for any other messages
that also pass over the affected link.

I hypothesize that flow-consistent routing is responsible for virtually all
of the congestion that occurs in the core of datacenter networks.

In-order packet delivery also causes hot spots in software.
For example, Linux performs load balancing in software by distributing
the handling of incoming packets across multiple cores; this is
essential in order to sustain high packet rates.
Each incoming packet is processed by the kernel
on two different cores before reaching the application (which may be
on a third core).
In order to ensure in-order packet delivery, all of the packets for a
given TCP connection must pass through the same sequence of cores.
This results in uneven core loading when two or more active connections
hash to the same core; again, the hot spot persists as long as the connections
are active. Measurements in \cite{homaLinux} indicate that hot spots are
the dominant cause of software-induced tail latency for TCP.

\emph{
Correction (1/2023): the first paragraph of this subsection is incorrect.
Out-of-order packet arrivals do not necessarily trigger packet retransmissions
in TCP.
Mechanisms such as triple-duplicate ACKs and RACK allow TCP to tolerate a
modest degree of packet reordering without retransmissions.
However, I am told by experts that asymmetries in datacenter networks
can cause significant packet reorderings that exceed TCP's tolerance.
In addition, performance optimizations in NICs and the Linux networking
stack, such as LRO and GRO, become ineffective with even modest reorderings,
resulting in significant performance degradation.
Thus, both networking hardware and Linux kernel software attempt
to preserve packet ordering, resulting in the problems described above.
}

\section{TCP is beyond repair}
\label{sec:beyondRepair}

One possible response to the problems with TCP is an incremental
approach, gradually fixing the issues while maintaining application
compatibility. There have already been numerous such attempts, and they
have made some progress. However, this approach is unlikely to succeed:
there are simply too many problems, and they are too deeply embedded in the
design of TCP.

As one example, consider congestion control. This aspect of TCP has
probably been studied more than any other in recent years, and
a number of novel and clever techniques have been devised.
One of the earliest was DCTCP~\cite{dctcp}; it provides significant
improvements in tail latency (see Figure~\ref{fig:p99_w4})
and has been widely implemented.
More recent proposals such as HPCC~\cite{hpcc} provide impressive additional
improvements (they are not included in Figure~\ref{fig:p99_w4} because
they don't have Linux kernel implementations).
However, all of these schemes are constrained by fundamental aspects
of TCP, such as its weak congestion signal based on buffer occupancy,
its inability to use switch priority queues, and its in-order
delivery requirement.
Significant
additional improvements will be possible only by breaking some of
TCP's fundamental assumptions.
The Homa curve in Figure~\ref{fig:p99_w4} shows that considerable
improvement is possible (though not shown in Figure~\ref{fig:p99_w4},
Homa also delivers better tail latency than newer proposals such as HPCC).

One of the problems with an incremental approach is that TCP has many
problems and they are interrelated. For example, the lack of message
boundaries makes it hard to implement SRPT and limits the amount of
information available for congestion control.
Thus, many different parts
of TCP will have to be changed before improvements will be visible.

In addition, the problems with TCP involve not just its implementation, but
also its API. In order to maximize performance in the datacenter, TCP would
have to switch from a model based on streams and connections to one based
on messages. This is a fundamental change that will affect applications.
Once applications are impacted, we might as well fix all of
the other TCP problems at the same time.

The bottom line is that there are no parts of TCP worth keeping. We need
a replacement protocol that is different from TCP in every significant aspect.
Fortunately, such a protocol already exists: Homa~\cite{homa, homaLinux}.
Homa provides an existence proof that all of TCP's problems are in fact
solvable.
\section{Homa}
\label{sec:homa}

Homa represents a clean-slate redesign of network transport for the
datacenter. Its design was informed by the problems with TCP as well as
experience using Infiniband~\cite{shanley2003infiniband} and
RDMA to implement large-scale datacenter applications.
Homa's design differs from TCP in
every one of the dimensions discussed in Section ~\ref{sec:tcpProblems}.
This section summarizes Homa's features briefly; for details, see
\cite{homa} and \cite{homaLinux}.

\subsection{Messages}
Homa is message-based. More precisely, it implements remote
procedure calls (RPCs), where a client sends a request message to a
server and eventually receives a response message. The primary
advantage of messages is that they expose dispatchable units to
the transport layer. This enables more efficient load balancing:
multiple threads can safely read from a single socket, and a NIC-based
implementation of the protocol could dispatch messages directly to a pool
of worker threads via kernel bypass. Having explicit message boundaries
also enables run-to-completion scheduling in the transport, such
as SRPT, and provides a more powerful congestion signal (see below).

Messages have one disadvantage relative to streams: it is difficult to
pipeline the implementation of a single large message. For example, an
application cannot receive any part of a message until the entire
message has been received. Thus a single large message will have higher
latency than the same data sent via a stream. However, large data
transfers can be handled by sending multiple messages in parallel, which
permits pipelining between messages.

\subsection{No connections}
Homa is connectionless.
There is no connection setup overhead, and an application
can use a single socket to manage any number of concurrent RPCs
with any number of peers.
Each RPC is handled independently: there are no ordering
guarantees between concurrent RPCs.

The state maintained by Homa falls into three major categories:
\begin{compactitem}
\item Sockets: Homa's state per socket is roughly equivalent to that for TCP,
but Homa applications can get by with a single socket, whereas TCP
applications require one socket per peer.
\item RPCs: Homa keeps about 300 bytes of state for each active RPC. This state
is discarded once the RPC has completed, so the total amount of state
is proportional to the number of active RPCs, not the total
number of peers.
\item Peers: each Homa host keeps about 200 bytes of state for each other
host, most of which is IP-level routing information.
This is much smaller than the
2000 bytes of state that TCP maintains per connection.
\end{compactitem}

In spite of its lack of connections, Homa ensures end-to-end reliability
for RPCs (or reports errors after unrecoverable network or host failures).
There is no need for applications to maintain additional timeouts.
Mechanisms such as flow control, retry, and congestion control are
implemented using per-RPC state;
one way of thinking about Homa is that it implements a short-lived and
lightweight connection for each RPC.

\subsection{SRPT}

Homa implements an SRPT scheduling policy in order to favor shorter
messages. It uses several techniques for this, of which the
most notable is that
it takes advantage of the priority queues provided by modern switches.
This allows higher-priority (shorter) messages to bypass packets queued for
lower-priority (longer) messages. As can be seen in Figure~\ref{fig:p99_w4}
this results in considerable improvements in tail latency compared to
either TCP or DCTCP. Messages of all lengths benefit from SRPT: even the
longest messages have significantly lower latency under Homa than under
TCP or DCTCP.

One potential concern about SRPT is that the longest messages might suffer
disproportionately high tail latencies or even starve.
This problem has not yet been observed in practice and is difficult to
produce even with an adversarial approach.
Nonetheless, the Linux implementation of
Homa contains an additional safeguard: a small fraction of each host's
bandwidth (typically 5--10\%) is dedicated to the \emph{oldest} message
rather than the smallest. This eliminates starvation and improves tail
latency for long messages in pathological cases, while still using
run-to-completion.

Homa's use of priority queues eliminates the ``pick your poison'' tradeoff
between latency and bandwidth discussed in Section~\ref{sec:tcpCongestion}.
Homa intentionally allows some buffers from longer messages to accumulate in
low-priority queues (\emph{overcommitment}); these ensure high link
utilization.
Short messages still achieve low latency since they use
higher priority queues.

\subsection{Receiver-driven congestion control}
Homa manages congestion from the receiver, not the sender. This makes
sense because the primary location for congestion is the receiver's downlink
(Homa eliminates core congestion as discussed in 
Section~\ref{sec:homaOutOfOrder} below).
The receiver has knowledge of all its incoming messages, so it is in a better
position to manage this congestion.
When a sender transmits a message, it can send a few \emph{unscheduled packets}
unilaterally (enough to cover the round-trip time), but the remaining
\emph{scheduled packets} may only be sent in response to \emph{grants}
from the receiver. With this mechanism, the receiver can limit congestion
at its downlink, and it also uses the grants to prioritize shorter messages.

Messages provide a powerful congestion signal that is not available in
stream-based protocols. Although message arrival is unpredictable, once
the first packet of the message has been seen, the total length of the
message is known. This enables proactive approaches
to congestion control, such as throttling other messages during this
message's lifetime and ramping them up again when this message
completes. In contrast, TCP can only be reactive, based on buffer
occupancy.

Incast can occur if many senders simultaneously send unscheduled packets,
but Homa's RPC orientation enables a simple mitigation;
see the Homa papers for details.

\subsection{Out-of-order packets}
\label{sec:homaOutOfOrder}
A key design feature of Homa is that it can tolerate out-of-order
packet arrivals.  This provides considerably more flexibility for load
balancing. For example, packet-level spraying can be used to distribute
packets across the network fabric instead of flow-consistent routing as in
TCP. If Homa becomes widely deployed, I hypothesize that
core congestion will cease to exist as a significant networking problem,
as long as the core is not systemically overloaded.
Homa's tolerance for out-of-order arrivals also allows
more flexibility for load balancing in software.

\subsection{Related work}
\label{sec:homaRelatedWork}
Several recent papers have claimed to identify problems with Homa and/or
to improve upon its performance, including Aeolus~\cite{aeolus},
PowerTCP~\cite{powertcp}, and DcPIM~\cite{dcpim}. However, all of
these papers have significant flaws, such as not implementing Homa
correctly or measuring it under artificial restrictions (e.g. Aeolus
uses statically buffer allocation in switches). For a more detailed
discussion of these papers, see the Homa Wiki~\cite{homaWiki}. The
Homa Wiki also contains a variety of other information about Homa,
and will be updated in the future to include new information and
related work.
\section{What about Infiniband?}
\label{sec:infiniband}

There are other TCP alternatives besides Homa, but none that
appear to meet the needs of datacenter computing.
One of the best known alternatives is Infiniband~\cite{shanley2003infiniband},
which has been widely adopted in the high-performance computing (HPC) arena,
and has recently seen increasing use in datacenters via RoCE, which
layers the RDMA API over Ethernet.

The primary advantage of RDMA is that it provides very low latency on
unloaded networks. It achieves this by offloading the transport protocol
implementation to the NIC and allowing user processes to bypass the
kernel and communicate directly with the NIC. Infiniband/RDMA NICs have
a well-deserved reputation for very high performance.

However, RDMA shares most of TCP's problems. It is based on
streams and connections (RDMA also offers unreliable datagrams, but
these have problems similar to those for UDP).
It requires in-order packet delivery.
Its congestion control mechanism, based on priority flow control (PFC),
is different from TCP's, but it is also problematic.
And, it does not implement an SRPT priority mechanism.

RDMA has the additional disadvantage that the NIC-based protocol
implementations are proprietary, so it is difficult to find out exactly how
they behave and to track down problems. As one example, the RAMCloud project
found several performance anomalies with Infiniband, especially
at high load; in most cases it was not possible to track them down because of
the closed nature of the implementation.

Future transport implementations should adopt Infiniband's kernel bypass
approach, but it seems unlikely that Infiniband itself can solve all
of TCP's problems.
\section{Getting there from here}
\label{sec:gettingThere}

It is hard to imagine a computing standard more entrenched than TCP,
so replacing it will be difficult. To make matters worse,
Homa (or any protocol that fixes all of TCP's problems) requires API
changes, meaning that application code will have to be modified.
Given the enormous number of applications that code directly to the
sockets interface, the task of modifying them all seems insurmountable,
at least for the near future.

Fortunately, it is not necessary to replace TCP for all applications.
The applications with the greatest need for a new transport protocol are
newer large-scale datacenter applications.
Most of these applications do not code to the sockets API; instead, they
are layered above one of a relatively small number of RPC frameworks such
as gRPC~\cite{grpc} or Apache Thrift~\cite{thrift}.
The easiest way to bring a new protocol into widespread use is to integrate
it with the major RPC frameworks.
This is a fairly manageable task, and once it
is done, applications using the frameworks can switch
to Homa with little or no work.

Work on framework integration has already begun: a gRPC driver for Homa
now exists for C++ applications, and Java support is underway.
This work is based on the Linux kernel implementation of Homa~\cite{homaLinux}.
\section{Conclusion}
\label{sec:conclusion}

TCP is the wrong protocol for datacenter computing. 
Every aspect of TCP's design is wrong: there is no part worth keeping.
If we want to eliminate the ``datacenter tax'', we must find a way to
move most datacenter traffic to a radically different protocol.
Homa offers an alternative that appears to solve all of TCP's problems.
The best way to bring Homa into widespread usage is integrate it with the
RPC frameworks that underly most large-scale datacenter applications.

\small
\bibliographystyle{abbrv}
\bibliography{local}




\end{document}